\documentstyle[pre,aps,epsf,multicol]{revtex}
\begin{document}

\title{Critical behavior in reaction-diffusion systems exhibiting absorbing
phase transition}
\author{G\'eza \'Odor}
\address{Research Institute for Technical Physics and Materials Science, \\
H-1525 Budapest, P.O.Box 49, Hungary}    
\maketitle

\begin{abstract}
Phase transitions of reaction-diffusion systems with site occupation
restriction and with particle creation that requires $n>1$ parents 
and where explicit diffusion of single particles ($A$) exists
are reviewed. 
Arguments based on mean-field approximation and simulations
are given which support novel kind of non-equilibrium criticality.
These are in contradiction with the implications of a suggested
phenomenological, multiplicative noise Langevin equation approach 
and with some of recent numerical analysis. 
Simulation results for the one and two dimensional 
binary spreading $2A\to 4A$, $4A\to 2A$ model display a new type of 
mean-field criticality characterized by $\alpha=1/3$ and $\beta=1/2$
critical exponents suggested in cond-mat/0210615.
\end{abstract}

\begin{multicols}{2}

\section{Introduction}
The classification of universality classes of second order phase transitions
is still one of the most important uncompleted task of statistical physics.
Recently phase transitions of genuine nonequilibrium systems have intensively 
been investigated among reaction-diffusion (RD) type of models exhibiting 
absorbing states \cite{Dick-Mar,Hin2000,dok}. 
There has been a hope that in such homogeneous systems symmetries and 
spatial dimensions are the most significant factors like in equilibrium 
ones, but gradually it turned out that there may be more relevant 
constituents as well. 
The most well known example is the parity conserving class (PC), 
which differs from the robust universality class of directed percolation (DP). 
The DP hypothesis stated by Janssen and Grassberger \cite{Jan81,Gras82}, 
according to which {\it in one component systems exhibiting continuous 
phase transitions to single absorbing state (without extra symmetry and 
inhomogeneity or disorder) short ranged interactions can generate DP class 
transition only}. However parity conservation itself proved to be an
insufficient condition in many cases \cite{Parkh,meod96,binary,OSC02}
and rather an underlying $A\to 3A$, $2A\to\emptyset$ (BARW2) process 
\cite{Cardy-Tauber} of particles 
and the $Z_2$ symmetry of absorbing states is necessary for this class 
\cite{nekimofcikk}.
On the other hand parity conservation in N-component branching and 
annihilating random walk (N-BARW) systems 
\cite{Cardy-Tauber}, or by triplet production models \cite{tripcikk}
was found to be responsible for novel classes again.
In one dimensional, multi-component reaction-diffusion systems
{\it site restriction} turned out to be a relevant, new factor 
\cite{barw2cikk,mexprocikk}.
Global conservation laws by directed percolation and lattice gas models
were shown to be irrelevant \cite{Tome,Sabag,Hil,Oliv}, while systems
with multiple absorbing states \cite{PCP} or with multi-components also 
exhibit DP class scaling behavior \cite{multipcpd}.

An other important puzzle was being investigated intensively 
during the past three years that emerges at phase transitions of 
binary production reaction-diffusion systems 
\cite{binary,OSC02,multipcpd,HT97,Carlon99,Hayepcpd,Odo00,HayeDP-ARW,coagcikk,NP01,MHUS,HH} (PCPD). 
In these systems particle production by pairs competes with pair 
annihilation and single particle diffusion. 
If production wins steady states with finite particle density appear
in (site restricted), while in unrestricted
(bosonic) models the density diverges. By lowering the 
production/annihilation rate a doublet of absorbing states without symmetries 
emerges. One of such states is completely empty, the other possesses a single 
wandering particle. In case of site restricted systems the transition to 
absorbing state is continuous. 
It is important to note that these models do not break the DP
hypothesis, because they exhibit multiple absorbing states which are not 
frozen, lonely particle(s) may diffuse in them. 
However {\it no corresponding symmetry or conservation law 
has been found yet}.
Non-DP type of phase transition in a binary production system was already 
mentioned in the early work of Grassberger \cite{GrasBP}. A corresponding
bosonic field theoretical model the annihilation fission (AF) process was
introduced and studied by Howard and T\"auber \cite{HT97}. These authors
claim a non-DP type of transition in AF, because the action does not 
contain linear mass term and the theory is non-renormalizible 
perturbatively unlike the Reggeon field theory of DP.
In field theories of models exhibiting DP class transition the 
canonical $A\to\emptyset$, $A\to 2A$ reactions
are generated by the renormalization transformation unlike here.
Further facts opposing DP criticality are the set of different 
mean-field exponents and the different upper critical dimension of binary
production PCPD like models ($d_c=2$ vs. $d_c=4$)\cite{HT97,OSC02}.

Forthcoming numerical studies reported somewhat different critical 
exponents, but there has been a consensus for about two years      
that this model should possess novel, non-DP type of transition.
The first density matrix study by Carlon et al. \cite{Carlon99} 
did not support a DP transition, but reported exponents near to
those of the PC class. Since the PCPD does not conserve the particle 
number modulo 2, neither exhibits $Z_2$ symmetric absorbing 
states the PC criticality was unfavored. Simulation studies by 
Hinrichsen \cite{Hayepcpd} and \'Odor \cite{Odo00} and 
coherent anomaly calculations by \'Odor \cite{Odo00,coagcikk} 
resulted in novel kind of critical behavior, although there
was an uncertainty in the precise values of critical exponents.
Exponent estimates showed diffusion ($D$) dependence that was 
under-pinned by pair mean-field results \cite{Odo00}, 
possessing two distinct classes as the function of $D$. 
Recently Park et al. reported well defined set of critical 
exponents in different versions of binary 
production PCPD-like processes \cite{PK66}. However these 
simulations were done at a fixed, high diffusion/annihilation rate
and agree with \'Odor's corresponding results \cite{pcpd2cikk}.
Kockelkoren and Chat\'e on the other hand claim an other set
of critical exponents \cite{KC0208497} that agrees with \'Odor's low 
diffusion/annihilation data. 

The PCPD model can be mapped onto a two-component model \cite{HayeDP-ARW} 
in which pairs are identified as a particle species following DP process 
and single particles as an other, coupled species following annihilating 
random walk. Simulations of such a two-component system at $D=0.5$
showed a continuous phase transition with exponents agreeing with
those of the PCPD for high diffusions. This model is similar to an other one 
\cite{woh}, which exhibits global particle number conservation as well.
Field theory \cite{woh} and simulations \cite{frei,ful} for the latter 
model reported two different universality classes as the function of $D$. 
It would be interesting to see if this conservation law is relevant or 
not like in case of the DP \cite{Oliv}.

Interestingly, higher level cluster mean-field approximations result in a 
single class behavior by varying $D$ and it turned out that by assuming 
logarithmic corrections the single class scenario can be supported by 
simulations too \cite{pcpd2cikk}. The origin of such logarithmic corrections 
may be a marginal perturbation between pairs and single particles 
in a coupled system description. A filed theoretical explanation
would be necessary. 

Two more recent studies \cite{DM0207720,HinDP1} reported non-universality 
in the dynamical behavior of the PCPD. While in the former one 
Dickman and Menezes explored different sectors 
(a reactive and a diffusive one) in the time evolution and gave 
non-DP exponent estimates, in the latter one Hinrichsen set afloat a
speculative conjecture that the ultimate long time behavior might be 
characterized by DP scaling behavior. 
In a forthcoming preprint \cite{HinDP2} Hinrichsen provided a discussion 
about the possibility of the DP transition based on a series 
of assumptions. His starting point is a Langevin equation that is
mapped onto a wetting process by Cole-Hopf 
transformation. By analyzing this process within a
certain potential he gave arguments for a DP transition. While this
Langevin equation with real noise is valid for the bosonic version 
of PCPD at and above the critical point, its usage in case of site occupancy 
restricted models is hypothetic, the noise can be complex 
at the transition point and may even change sign by the
transformation. 
Furthermore the diffusive field of solitary particles is neglected.

In a very recent preprint \cite{BarkCar} Barkema and Carlon continue
this line and show that some simulation and density matrix 
renormalization results may also be interpreted as a signal of a
phase transition belonging to the DP class. 
By assuming correction to scaling exponents that are equal 
to DP exponents and relevant up to quadratic or 3-rd order in
the asymptotic limit they fitted their numerical results in case 
of two independent exponents. The extrapolations resulted is 
close to DP values for $D=0.5$. 
However for smaller $D$-s and by surface critical exponents this 
technique gave exponent estimates which are out of the error margin of DP. 
 
An other novel class that may appear in triplet production systems was 
proposed in \cite{KC0208497,PHK02} (TCPD). This reaction-diffusion model 
differs from the PCPD that for a new particle generation at least three 
particles have to meet. For such generalizations Park et al. proposed 
a phenomenological Langevin equation that exhibits real, multiplicative 
noise \cite{PHK02}. By simple power counting they found that the triplet 
model exhibits distinct mean-field exponents and upper critical 
dimension $4/3\le d_c \le 8/3$.
The simulations in 1d \cite{PHK02} indeed showed non-trivial
critical exponents, which do not seem to correspond to any known 
universality classes. 
Kockelkoren and Chat\'e reported similar results in stochastic cellular 
automata (SCA) versions of general $nA\to (n+k)A$, $mA\to(m-l)A$ 
type of models \cite{KC0208497}, where multiple particle creation on 
a given site is suppressed by an exponentially decreasing creation 
probability ($p^{N/2}$) of the particle number.
They claim that their simulation results in 1d are in agreement with the
fully occupation number restriction counterparts and set up a general table 
of universality classes, where as the function of $n$ and $m$ only 4 
non-mean-field classes exist, namely the DP class, the PC class, 
the PCPD and TCPD classes. However more extensive simulations of 1 and 2 
dimensional site restricted lattice models \cite{tripcikk} do not 
support some of these results in case of different triplet and 
quadruplet models. In the $3A\to 4A$ $3A\to\emptyset$ triplet
model 1d numerical data can be interpreted as mean-field behavior 
with logarithmic corrections and in two dimensions
clear mean-field exponents appear, hence the upper critical
dimension is $d_c=1$, which contradicts the Langevin equation prediction.
Surprisingly other non-trivial critical behavior were also detected
in the $3A\to 5A$ $3A\to A$ parity conserving triplet model and in some 
quadruplet models \cite{tripcikk}. The cause of 
differences between the results of these studies is subject of 
further investigations. Again proper field theoretical treatment
would be important.

The classification of universality classes of nonequilibrium
systems by the exponent $\mu$ of a multiplicative noise in the Langevin 
equation was suggested some time ago by Grinstein et al. \cite{GMTMN}.
However it turned out that there may not be corresponding particle systems
to real multiplicative noise cases \cite{HT97} and an imaginary part 
appears as well if one derives the Langevin equation of a RD system 
starting from the master equation in a proper way. Furthermore for 
higher-order processes the emerging nonlinearities in the
master equation action do not allow a rewriting in terms of
Langevin-type stochastic equations of motion, hence for high-order 
processes like those of the TCPD a Langevin representation may not
exist.

This situation resembles to some extent to a decade long debate over the 
critical phase transition of driven diffusive systems \cite{Val,Leu,ZSS}.
The latest papers in this topic suggest that the phenomenological 
Langevin equation originally set up for such systems do not correspond 
exactly to the lattice models investigated. Simulations for different
lattice models show, that instead of an external current the anisotropy is the
real cause of the critical behavior observed in simulations 
\cite{IDLG,Alb02}.

\section{Mean-field classes}

In this section I show that mean-field classes of site restricted 
lattice models with general microscopic processes of the form
\begin{equation}
n A \stackrel{\sigma}{\to} (n+k)A, 
\qquad m A \stackrel{\lambda}{\to} (m-l) A, \label{genreactions}
\end{equation}
with $n>1$, $m>1$, $k>0$, $l>0$ and $m-l\ge 0$ are different from
those of the DP and PC processes backing numerical results which
claim novel type of criticality below $d_c$.
The mean-field equation that can be set up for the lattice version
of these processes (with creation probability $\sigma$ and
annihilation or coagulation probability $\lambda=1-\sigma$) is
\begin{equation}
\frac {\partial\rho}{\partial t}= 
a k \sigma \rho^n (1-\rho)^k - a l (1-\sigma) \rho^m, \label{MFeq}
\end{equation}
where $\rho$ denotes the site occupancy probability and
$a$ is a dimension dependent coordination number.
Each empty site has a probability (1-$\rho$) in mean-field approximation,
hence the need for $k$ empty sites at a creation brings in a $(1-\rho)^k$
probability factor. By expanding $(1-\rho)^k$ and keeping the lowest 
order contribution one can see that for site restricted lattice 
systems a $\rho^{n+1}$-th order term appears with negative 
coefficient that regulates eq.(\ref{MFeq}) in the active phase. 
In the inactive phase one expects a dynamical behavior dominated by the
$mA\to\emptyset$ process, for which the particle density decay law is
known $\rho(t)\propto t^{1/(m-1)}$ \cite{Cardy-Tauber}.
The steady state solutions were determined in \cite{tripcikk} 
analytically and one can distinguish three different situations
at the phase transition: (a) $n=m$, (b) $n>m$ and (c) $n<m$ .

\subsection{The $n=m$ symmetric case}

As discussed in \cite{tripcikk} the leading order singularities
of steady state solution can be obtained. By approaching the
the critical point $\sigma_c=\frac{l}{k+l}$ in the active phase
the steady state density vanishes continuously as 
\begin{equation}
\rho \propto |\sigma-\sigma_c|^{\beta^{MF}},
\end{equation}
with the order parameter exponent exponent $\beta^{MF}=1$. 
At the critical point the density decays with a power-law
\begin{equation}
\rho \propto t^{-\alpha_{MF}} \ ,
\end{equation}
with $\alpha_{MF}=\beta^{MF}/\nu^{MF}_{||}=1/n$, hence 
$\nu^{MF}_{||}=n$, providing a series of mean-field
universality classes for $n>1$ 
(besides DP an PC where $\nu^{MF}_{||}=1$)
and backing the results, which claim novel type of non-trivial 
transitions below the critical dimension.
Unfortunately determining the the value of $d_c$ is a 
non-trivial task without a proper Langevin equation.
These scaling exponents can be obtained from bosonic, coarse grained 
formulation too \cite{PHK02}, where a $\rho^{n+1}$-th order term, 
with negative coefficient had to be added by hand to suppress 
multiple site occupancy. 
It is known however that hard-core particle blocking may result in 
relevant perturbation in $d=1$ dimension \cite{mexprocikk}, so for 
cases where the upper critical dimension is $d_c\ge 1$ the
site restricted, $N>1$ cluster mean-field approximation that takes
into account diffusion would be a more adequate description of
the model (see \cite{parwcikk}).

\subsection{The $n>m$ case} \label{MFnmsect}

In this case the mean-field solution provides first order
transition (see \cite{tripcikk}), hence it does not 
imply anything with respect to possible classes for models below 
the critical dimension ($d<d_c$). Note however, that by higher order 
cluster mean-field approximations, where the diffusion plays 
a role the transition may turn into continuous one 
(see for example \cite{boccikk,OdSzo,meorcikk}).
The simulations by Kockelkoren and Chat\'e report DP class transition
for such models in one dimension \cite{KC0208497}.

\subsection{The $n<m$ case} 

In this case the critical point is at zero branching rate $\sigma_c=0$,
where the density decays with $\alpha^{MF}=1/(m-1)$ as in case of the 
$n=1$ branching and $m=l$ annihilating models showed by 
Cardy and T\"auber \cite{Cardy-Tauber} (BkARW classes).
However the steady state solution for particle production with $n>1$ 
parents gives different $\beta$ exponents than those of BkARW classes, 
namely $\beta^{MF}=1/(m-n)$, defining a whole new series of 
mean-field classes \cite{tripcikk}, for a simplicity I shall call them 
PkARW classes. 
It is important to note, that one has not found a corresponding symmetry 
or conservation law to these classes. 
These mean-field classes imply novel kind of critical behavior for $d<d_c$.
For $n=2$, $m=3$ the mean-field exponents are $\beta=1$ and $\alpha=1/2$
agreeing with the mean-field exponents of the PCPD class.
Indeed for $n=2$, $m=3$ \cite{KC0208497} reports PCPD class dynamical
criticality. This supports the expectation that non-mean-field classes 
follow the distinctions observed in the corresponding mean-field classes. 
To go further in sections \ref{2dsimu} and \ref{1dsimu} I investigate 
the phase transition of the simplest unexplored PkARW classes, in
the $n=2$, $m=4$ model.

\subsection{The role of $k$ and $l$}

In the mean-field approximation $k$ and $l$ do not affect the universal
properties, however simulations in one dimension showed \cite{tripcikk}
that in case of the $m=n=k=l=3$ model the critical point was shifted to 
zero branching rate and a BkARW class transition emerged there contrary to 
what was expected for $n=m$. 
For a stochastic cellular automaton version of these reactions 
\cite{KC0208497} reported a non-trivial critical transition. 
In general one may expect such effects for large $k$ and $l$ values,
for which N-cluster mean-field approximation -- that takes into
account diffusion -- would give a better description.

\section{Simulations of the $2A\to 4A$, $4A\to 2A$ model 
in two dimensions} \label{2dsimu}

In the \ref{MFnmsect} section I introduced PkARW mean-field classes 
for $n<m$. 
Here I explore the phase transition in the simplest model from this 
class, in the $2A\to 4A$, $4A\to 2A$ model with $D=0.5$ diffusion rate.
Two dimensional simulations were performed on $L=1000$ linear 
sized lattices with periodic boundary conditions. One Monte Carlo step (MCS)
--- corresponding to $dt=1/P$ (where $P$ is the number of particles) ---
is built up from the following processes. A particle and a number 
$x \in (0,1)$ are selected randomly; if $x < D =0.5$ a site exchange 
is attempted with one of the randomly selected empty nearest neighbors (nn); 
if $x\ge D = 0.5$ two particles are created with probability $\sigma$
at randomly selected empty nn sites provided the number of nn particles was
greater than or equal $2$; or if $x\ge 0.5$ two particles are 
removed with probability $1-\sigma$.  
The simulations were started from fully occupied lattices and
the particle density ($\rho(t)$) decay was followed up to 
$4\times 10^5$ MCS.

\begin{figure}
\epsfxsize=70mm
\epsffile{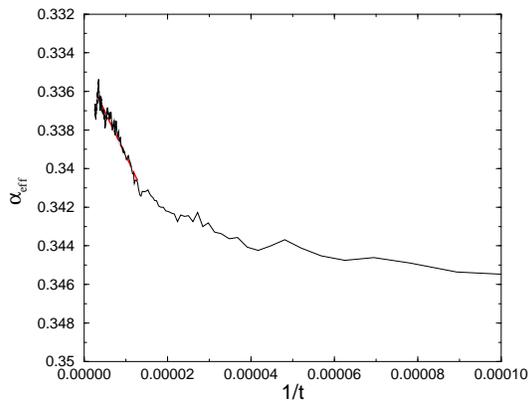}
\vspace{4mm}
\caption{$\alpha_{eff}(1/t)$ in the two dimensional
$2A\to 4A$, $4A\to 2A$ model at $\sigma_c=0$. The dashed line shows a
linear fitting for $t>10^5$ MCS resulting in $\alpha=0.334(1)$.}
\label{2A4A}
\end{figure}

First the critical point was located by measuring the dynamic behavior
of $\rho(t)$. It turned out that the transition is at zero branching
rate ($\sigma_c=0$). The density decay was analyzed by the local slopes 
defined as
\begin{equation}
\alpha_{eff}(t) = {- \ln \left[ \rho(t) / \rho(t/m) \right] 
\over \ln(m)} \label{slopes}
\end{equation}
where I used $m=4$. As Fig.\ref{2A4A} shows the local slopes
curve for $t>10^5$ MCS extrapolates to the mean-field value 
$\alpha=0.334(1)$.
This value agrees with the mean-field value $\alpha^{MF}=1/3$.
Density decays for several $\sigma$-s in the active phase 
($0.0002 \le \sigma \le 0.05$) were followed on logarithmic time 
scales and averaging was done over $\sim 100$ independent runs 
in a time window, which exceeds the level-off time by a decade.
The steady state density in the active phase near the critical 
phase transition point is expected to scale as
\begin{equation}
\rho(\infty,\sigma) \propto |\sigma-\sigma_c|^{\beta} \ .
\end{equation}
\begin{figure}
\begin{center}
\epsfxsize=70mm
\centerline{\epsffile{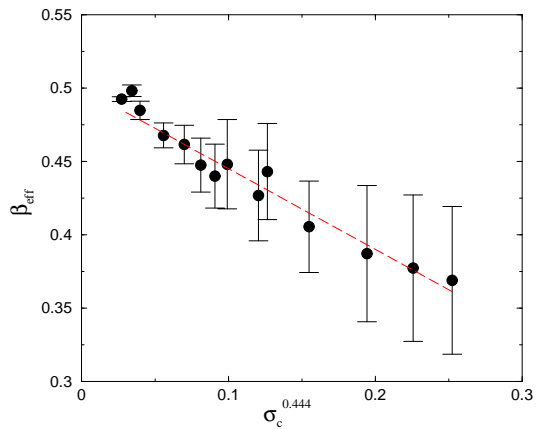}}
\caption{$\beta_{eff}$ as the function of $\sigma^{-\delta}$ in the 
two dimensional $2A\to 4A$, $4A\to 2A$ model (bullets). 
The dashed line shows a fitting of the form (\ref{corsca}). }
\label{beta2A4A}
\end{center}
\end{figure}
Using the local slopes method one can get a precise estimate for
$\beta$ as well as for the corrections to scaling
\begin{equation}
\beta_{eff}(\sigma_i) = \frac {\ln \rho(\infty,\sigma_i) -
\ln \rho(\infty,\sigma_{i-1})} {\ln(\sigma_i) - \ln(\sigma_{i-1})} \ .
\label{beff}
\end{equation} 
As one can see on Fig.\ref{beta2A4A} the effective exponent clearly tends
to the expected mean-field value $\beta=0.5$ as $\sigma\to 0$.
Assuming a correction to scaling of the form 
\begin{equation}
\beta_{eff}=\beta - a t^{-\delta} \label{corsca}
\end{equation} 
non-linear fitting results in $\delta=0.44(1)$ correction to scaling
exponent. 

Besides these scaling correction assumptions I also tried to apply different, 
lowest order logarithmic corrections to the data, but these fittings
gave exponents slightly away from mean-field values and the 
corresponding coefficients proved to be very small, therefore I concluded 
that $d_c<2$. In the next section I do the same analysis in $d=1$.

\section{Simulations of the $2A\to 4A$, $4A\to 2A$ model 
in one dimensions} \label{1dsimu}

\begin{figure}
\epsfxsize=70mm
\epsffile{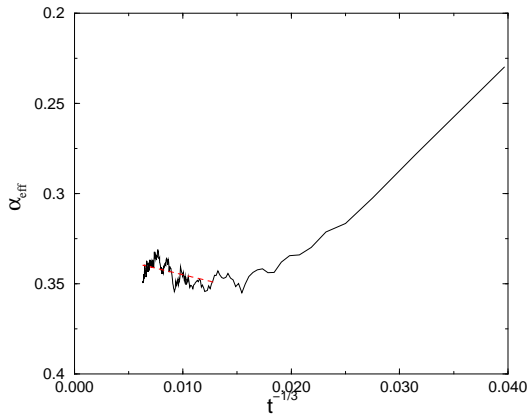}
\vspace{4mm}
\caption{$\alpha_{eff}(t^{-1/3})$ in the one dimensional
$2A\to 4A$, $4A\to 2A$ model at $\sigma_c=0$. The dashed line shows a
linear fitting for $t> 5\times 10^5$ MCS resulting in $\alpha=0.32(2)$.}
\label{2A4A_5}
\end{figure}

\begin{figure}
\begin{center}
\epsfxsize=70mm
\centerline{\epsffile{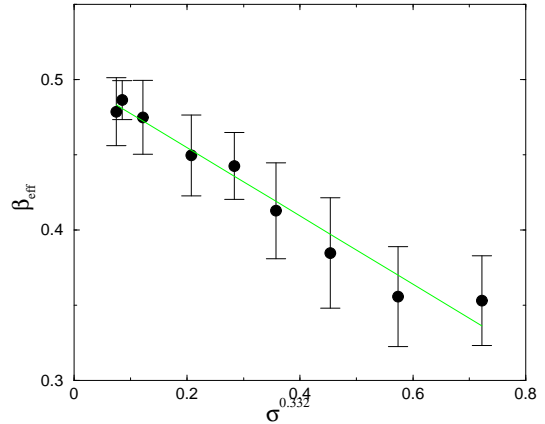}}
\caption{$\beta_{eff}$ as the function of $\sigma^{-\delta}$ in the 
one dimensional $2A\to 4A$, $4A\to 2A$ model (bullets). 
The dashed line shows a fitting of the form (\ref{corsca}). }
\label{beta2A4A_5}
\end{center}
\end{figure}

The simulations in one dimension were carried out on $L=20000$ sized
systems with periodic boundary conditions. The initial states were again
fully occupied lattices, and the density of particles is followed up to
$4\times 10^6$ MCS. An elementary MCS consists of the following processes:
\begin{description}
\item[(a)] $A\emptyset\leftrightarrow\emptyset A$ with probability D,
\item[(b)] $AAAA \to \emptyset AA\emptyset$ with probability $(1-\sigma)(1-D)$,
\item[(c)] $AA\emptyset\emptyset\to AAAA$ or $\emptyset\emptyset AA\to AAAA$
with probability $\sigma (1-D)$,
\end{description}
The critical point was located at $\sigma_c=0$ again. As one can see on
Figure \ref{2A4A_5} there is a crossover of the local slopes for 
$t > 5\times 10^5$ MCS and a linear extrapolation for this region results
in $\alpha=0.329(5)$ agreeing with the mean-field value.
The steady state data were analyzed in the active region for
$0.0003 \le \sigma \le 0.5$ as in two dimensions, by the local 
slopes method (eq. \ref{beff}) and by assuming correction to scaling
of the form (\ref{corsca}). This resulted in $\delta'=0.332$ correction to
scaling exponent. The local slopes plotted as the function 
$\sigma^{-\delta}$ shown on Fig. \ref{beta2A4A_5} extrapolates to
$\beta=0.49(1)$ in agreement with the mean-field exponent.

\section{Conclusions}

In this paper I reviewed and discussed the state of the art of
the phase transitions of reaction-diffusion systems exhibiting 
explicit diffusion and production by $n>1$ parents.
Arguments are given against DP criticality that has recently been
suggested in some papers.  These are supported by a series of mean-field 
classes that can be classified by the existence of a $n=m$ symmetry 
in the system and by the $n$ and $m$ values.
Especially the need for a proper field theoretical treatment is emphasized.
The upper critical dimension in these models is not known. 

I determined by simulations the $\alpha$ and $\beta$ exponents of the 
$2A\to 4A$, $4A\to 2A$ model in one an two dimensions.
These results indicate that for this binary production system
the upper critical dimension is $d_c < 1$.
This model conserves the parity of particles still its transition does not 
belong to the PC class.
Note that while the density decay results in one dimension are
in agreement with that of Kockelkoren and Chat\'e-s simulations 
\cite{KC0208497}, the off-critical order parameter exponent 
is $\beta^{MF}=1/(m-n)$ which shows that there are more classes 
exist at zero branching rate besides the BkARW universality classes. 
It is still an open question if there is any variant of PkARW models 
that exhibits non-mean-field criticality in physical dimensions ($d\ge 1$).
\bigskip

{\bf Acknowledgements:}\\

Support from Hungarian research funds OTKA (Grant No. T-25286)
is acknowledged. The author thanks the trial access to the 
NIIFI Cluster-GRID of Hungary.

\end{multicols}
\end{document}